\documentclass[12pt]{iopart}

\usepackage{tabularx}
\usepackage{booktabs}
\usepackage{caption}
\usepackage{subcaption}
\usepackage{graphicx}

\expandafter\let\csname equation*\endcsname=\relax 
\expandafter\let\csname endequation*\endcsname=\relax 

\usepackage{slashed}
\usepackage{mathtools}
\usepackage{amssymb}
\usepackage{amsmath}
\usepackage{mathrsfs}
\usepackage{physics}
\usepackage{bm}
\usepackage{gensymb}

\usepackage[backend=biber,
            style=numeric-comp,
            sorting=none,
            doi=true,
            url=true,
            eprint=true]{biblatex}
\usepackage[table,xcdraw]{xcolor}
\usepackage{hyperref}

\definecolor{cite}{rgb}{0.,0.,0.9}
\hypersetup{colorlinks,linkcolor=black,citecolor={cite},urlcolor={cite}}

\DeclarePairedDelimiter\avg{\langle}{\rangle}

\makeatletter
\DeclareRobustCommand{\cev}[1]{%
  {\mathpalette\do@cev{#1}}%
}
\newcommand{\do@cev}[2]{%
  \vbox{\offinterlineskip
    \sbox\z@{$\m@th#1 x$}%
    \ialign{##\cr
      \hidewidth\reflectbox{$\m@th#1\vec{}\mkern4mu$}\hidewidth\cr
      \noalign{\kern-\ht\z@}
      $\m@th#1#2$\cr
    }%
  }%
}
\makeatother

\eqnobysec

\makeatletter
\newcommand{\mainmatter}{%
  \setcounter{footnote}{0}%
  \patchcmd{\@makefntext}{\fnsymbol}{\arabic}{}{}%
  \patchcmd{\@thefnmark}{\fnsymbol}{\arabic}{}{}%
  \def\@makefnmark{\textsuperscript{\arabic{footnote}}}%
}
\makeatother

\begin{document}

\title{Gravitational waves decohere quantum superpositions}

\author{Flynn Linton$^1$\footnote{Corresponding author.} and Shubhanshu Tiwari$^2$}
\address{$^1$Department of Physics, ETH Z\"{u}rich}
\address{$^2$Physics Institute, University of Z\"{u}rich}
\ead{\mailto{flynn.linton@monash.edu}, \mailto{shubhanshu.tiwari@physik.uzh.ch}}


\begin{abstract}
Understanding the interplay between quantum mechanical systems and gravity is a crucial step towards unifying these two fundamental ideas. Recent theoretical developments have explored how global properties of spacetime would cause a quantum spatial superposition to lose coherence. In particular, this loss of coherence is closely related to the memory effect, which is a prominent feature of gravitational radiation. In this work, we explore how a burst of gravitational radiation from a far-away source would decohere a quantum superposition. We identify the individual contributions to the decoherence from the memory and oscillatory components of the gravitational wave source, corresponding to soft and hard graviton emissions, respectively. In general, the memory contributions dominate, while the oscillatory component of the decoherence is strongly dependent on the phase of the burst when it is switched off. This work demonstrates how quantum systems can lose coherence from interactions with a classical gravitational field. We also comment on the electromagnetic analogue of this effect and discuss its correspondence to the gravitational case.



\end{abstract}

\vspace{2pc}

\maketitle


\mainmatter


\section{Introduction}
One of the greatest open problems in physics today is to understand whether gravity is quantum in nature. An important step in potentially unifying these ideas is to understand the semi-classical limit in which quantum mechanical systems and classical gravity interact. A recent series of papers \cite{danielson_gravitationally_2022, danielson-decoherence1, danielson-decoherence2, local-decoherence, Danielson:2025iud, Gralla:2023oya, Wilson-Gerow:2024ljx} based on a gedankenexperiment, originally posed by Mari \textit{et al} \cite{mari_experiments_2016}, has cleverly demonstrated how black holes, as well as the local and global properties of a spacetime, can effect an experimenter’s ability to maintain a coherent quantum spatial superposition.

The setup of the gedankenexperiment is as follows. An experimenter, Alice, begins with a quantum state that she sends through a Stern-Gerlach device over time $T_1$ to create a spatially separated superposition, whose components are separated by $d_0$. Her superposition travels along an inertial trajectory for the duration $T$, after which she recombines her state using a reverse Stern-Gerlach device over time $T_2$. The components of Alice's state produce a superimposed gravitational field that can penetrate the Killing horizon of a spacetime. If an observer, Bob, sits behind the horizon, he cannot causally influence Alice's experiment. Bob may acquire `which-path’ information about Alice's superposition by measuring her gravitational field, thereby entangling himself with her field and causing her state to lose coherence. If Bob can perform his measurement sufficiently quickly before Alice recombines her superposition, then a paradox arises in which one of causality or complementarity must be violated. If causality holds, Alice's superposition, which is already in a maximally entangled Bell state, must become entangled with Bob's particle, thereby violating complementarity\footnote{By complementarity, we refer to `conservation of entanglement,' in the sense that Alice's state cannot become more entangled if it is already a maximally entangled state.}. Conversely, if complementarity holds and Alice and Bob's states remain unentangled with one another, faster-than-light information must be transmitted between the two observers due to their spacelike separation, violating causality.

To resolve this supposed paradox, Danielson \textit{et al} \cite{danielson_gravitationally_2022} concluded that it is the interaction between internal degrees of freedom in Alice's system that cause it to be disturbed and thus decohere. A quantitative estimate of the minimum amount Alice's state decoheres during her experiment was formulated in \cite{danielson-decoherence1} for spacetimes containing a black hole. This notion of decoherence was later generalised to spacetimes that contain Killing horizons \cite{danielson-decoherence2} and bifurcate Killing horizons \cite{Gralla:2023oya}. Local formulations of the decoherence were then developed by quantising Alice's gravitational field and defining the two-point function at asymptotic times \cite{local-decoherence} and as the action of a thermal bath on her experiment \cite{Wilson-Gerow:2024ljx}.

Danielson \textit{et al} \cite{danielson_gravitationally_2022, danielson-decoherence1, local-decoherence} showed that the decoherence is directly related to the memory effect, which is a prominent feature of gravitational radiation. In this paper, we study the decohering effect of gravitational radiation on quantum spatial superpositions. We extend the gedankenexperiment described above such that Alice's state is subject to a short burst of gravitational waves that lasts for $T_{\text{GW}} \ll T$ while she maintains her superposition. In particular, we suppose Alice sits in an approximately Minkowski spacetime with a distant gravitational wave source that can be switched on and off. The gravitational waves from the source thus satisfy the free linearised Einstein equations, whose solutions are given by plane waves. The setup of our modified gedankenexperiment is shown in Figure \ref{fig:modified-gedankenexperiment}.

The remainder of this paper is structured as follows. In section \ref{ch:local-decoherence}, we review the construction of a local description of Alice's decoherence originally formulated by Danielson \textit{et al} \cite{local-decoherence}. Furthermore, we review their calculation of the decoherence when Alice carries out her experiment in a purely Minkowski spacetime. The details of this calculation will be necessary in extending these ideas to the case where we have a gravitational wave source present. In section \ref{ch:decoherence_gws}, we introduce a distant gravitational wave source while Alice performs her experiment and determine her decoherence using (i) the classical gravitational quadrupole and (ii) the local decoherence formalism. We then identify the memory and oscillatory components of the gravitational wave as contributing to soft and hard graviton emission, respectively, and quantify the decoherence due to each of these.

\section{Local description of decoherence} \label{ch:local-decoherence}
In this section, we review the construction of a local description of Alice's decoherence and perform the calculation in Minkowski spacetime, originally done by Danielson \textit{et al} \cite{local-decoherence}.

The degree of decoherence of Alice's state can be determined by constructing the initial and final states of her particle. After creating her spatially separated superposition using a Stern-Gerlach device, the state of Alice's particle is
\begin{equation}
    \frac{1}{\sqrt{2}} \left( \ket*{\uparrow ; A_1} \ket*{\psi_1} + \ket*{\downarrow ; A_2} \ket*{\psi_2} \right), \label{eq:Alice-total-state}
\end{equation}
where $\ket*{\uparrow}$ and $\ket*{\downarrow}$ are eigenstates of the $z$-spin operator, $\ket*{A_1}$ and $\ket*{A_2}$ are spatially separated wavepackets and $\ket{\psi_i}$ are coherent states of the gravitational field generated by each component of her superposition. At asymptotically late times, the field naturally decomposes into its radiative and static components \cite{local-decoherence}. The static (Newtonian) part of the field is not an independent degree of freedom since it moves with the components of the superposition to future timelike infinity. Therefore, at asymptotically late times (i.e. after she recombines her superposition) her state is
\begin{equation}
    \frac{1}{\sqrt{2}} \left( \ket*{\uparrow ; A_1}_{i^+} \ket*{\Psi_1}_{\mathscr{I}^+} + \ket*{\downarrow ; A_2}_{i^+} \ket*{\Psi_2}_{\mathscr{I}^+} \right), \label{eq:late-time-state}
\end{equation}
where $\ket{\Psi_i}$ is the \textit{radiative} part of the gravitational state $\ket{\psi_i}$ which `which-path' information can be extracted from by an observer. Here, $i^+$ and $\mathscr{I}^+$ denote future timelike infinity and future null infinity, respectively. 

After recombination, the spatial components of Alice's superposition coincide, so $\ket{A_1} = \ket{A_2}$. The degree to which the radiative states differ depends on the off-diagonal elements of the density matrix describing the radiation state of \eqref{eq:late-time-state} and defines the decoherence of Alice's particle,
\begin{equation}
    \mathscr{D} = 1-\left| \braket{\Psi_1}{\Psi_2}_{\mathscr{I}^+} \right|. \label{eq:decoherence-alice-nullinf}
\end{equation}
Single-particle states of the gravitational field of each component of Alice's superposition are coherent quantum states with inner product \cite{danielson-decoherence2}
\begin{equation}
    |\braket{\Psi_1}{\Psi_2}| = \exp(-\frac{1}{2} \avg{N}), \label{eq:coherent-inner-product}
\end{equation}
where $\avg{N}$ is the number of gravitons Alice radiates to null infinity. The decoherence of Alice's state is therefore
\begin{equation}
    \mathscr{D} = 1-\exp(-\frac{1}{2} \avg{N}).
\end{equation}
For $\avg{N} \gtrsim 1$, $\mathscr{D} \sim 1$ so her state decoheres. If Alice is able to recombine her particle adiabatically, such that she does not emit any gravitational radiation, then any difference between her gravitational field state before creating the superposition and after its recombination would be a case of Unruh’s false loss of coherence \cite{Unruh:1999vn}. Furthermore, if there is no observer to harvest the gravitational radiation she emits, no `which-path' information is obtained from her superposition and any difference in her state would correspond to a false loss of coherence.

Alice can, in principle, reduce the `actual' decoherence to zero by taking an infinitely long time to create and recombine her state (i.e. in the limit as $T_1, T_2 \to \infty$), ensuring that she emits no gravitational radiation and provided she can maintain the superposition for an arbitrarily long time $T$.

We can estimate the decoherence by determining the number of gravitons Alice's state emits during her experiment from the two-point function of her gravitational field, which depends on the difference in stress-energy $T_1^{ab} - T_2^{ab}$ between each component of the superposition. Provided the components of her state are well localised, their stress-energy can be treated as point-like particles. Further assuming minimal self-interactions, Danielson \textit{et al} \cite{local-decoherence} showed the number of gravitons Alice emits is
\begin{equation}
    \avg{N} = m^2 \int \rmd t \, \rmd t' \, d^2(t) d^2(t') \avg{s^a s^b E^{\text{in}}_{ab}(t,\bm{X})s^c s^d E^{\text{in}}_{cd}(t',\bm{X})}_{\Omega}. \label{eq:N-2}
\end{equation}
Here, $E_{ab}^{\text{in}} = C_{acbd} t^c t^d$ is the quantum field observable describing the electric part of the Weyl tensor on static time-slices of the spacetime whose vacuum state is $\ket{\Omega}$ and $\bm{X}(t)$ is the centre-of-mass position of Alice's lab. The unit vector $s^a$ points in the direction of separation between the components of the superposition. We assume that $s^a$ is Lie transported along $t^a$ such that the direction of separation is constant in time.

It was pointed out in \cite{local-decoherence} that since determining $\avg{N}$ amounts to calculating the two-point function of Alice's unperturbed gravitational field, the decoherence can be determined independently of the particle content in her late-time field state. In particular, the decoherence is a purely local phenomenon due to vacuum fluctuations within Alice's lab.

Zhou \& Yu \cite{Zhou_2012} and Menezes \cite{Menezes2016} calculated this quantity in the context of electromagnetism for radially separated states, to which they found the two-point function of the electric field given a generic vacuum state. In general relativity, the two-point function of the electric part of the Weyl tensor was found similarly \cite{local-decoherence}. In Minkowski spacetime,
\begin{equation}
    \avg{N} \sim m^2 \sum_{l=2}^{\infty} \frac{1}{r^6} \int_{0}^{\infty} \frac{\rmd \omega}{\omega} |\hat{d^2}(\omega)|^2 |\cev{R}_{\omega l}(r)|^2, \label{eq:N-Minkowski}
\end{equation}
where
\begin{equation}
    \hat{d^2}(\omega) = \int_{-\infty}^{\infty} \rmd t \, d^2(t) \rme ^{\rmi\omega t}. \label{eq:d(t)-FT}
\end{equation}
The quantities $\cev{R}_{\omega l}$ are gravitational modes defined by the spherical harmonic decomposition of gravitational wave solutions in a spherically symmetric and static spacetime. In particular, the purely radial physical solutions are
\begin{equation}
    h_{rr}^{(n ; \omega l m)}(r,\theta,\phi) = R^{(n)}_{\omega l}(r) Y_{l m}(\theta,\phi) \rme ^{-\rmi \omega t},
\end{equation}
which follows analogously from the quantisation of the electromagnetic field in such a spacetime done by Crispino \textit{et al} \cite{crispino_quantization_2001}. Here, $Y_{lm}$ is a spherical harmonic, with $l \geq 2$. The index $n$ distinguishes modes incoming from past null infinity $\mathscr{I}^-$ ($n=\leftarrow$) and white hole modes incoming from the past Cauchy horizon $H^-$ ($n=\rightarrow$).

The incoming modes from $\mathscr{I}^-$ were calculated for Schwarzschild spacetimes\footnote{See (4.31) in \cite{local-decoherence}.} in \cite{local-decoherence}, from which the Minkowski mode functions are found by taking the limit as the Schwarzschild mass vanishes. Namely,
\begin{equation}
    \cev{R}_{\omega l}(r) = -2i^{3l+1} \omega r j_l(\omega r), \label{eq:mode-function}
\end{equation}
where $j_l$ is a spherical Bessel function. Before constructing her superposition and after recombining her state, the separation between the components of her superposition is effectively zero, so the Fourier transform \eqref{eq:d(t)-FT} is
\begin{equation}
    \hat{d^2}(\omega) = \int_{-T/2-T_1}^{T/2+T_2} \rmd t \, \rme ^{\rmi\omega t} d^2(t).
\end{equation}
The creation and recombination of Alice's superposition contributes negligibly to the Fourier transform for $T \gg T_1,T_2$ \cite{local-decoherence}, so we can approximate
\begin{equation}
    \hat{d^2}(\omega) \approx \int_{-T/2}^{T/2} \rmd t \, \rme ^{\rmi\omega t} d^2(t) = \frac{2 d_0^2 \sin(\omega T/2)}{\omega},
\end{equation}
Here, we have used that the components of Alice's superposition remain at a constant separation $d(t)=d_0$ for $t \in (-T/2,T/2)$. The oscillatory term is bounded, so the Fourier transform is proportional to $1/\omega$, indicating soft modes ($\omega \ll 1$) will dominate the integral in $\avg{N}$. As $\omega \to 0$, $\hat{d^2}(\omega)$ diverges. For the decoherence to remain finite, it is therefore necessary to introduce an infrared frequency cutoff $\omega_{\text{IR}} \sim 1/T$, which describes the minimum energy of a graviton Alice can radiate during her experiment. We also introduce an ultraviolet cutoff $\omega_{\text{UV}} \sim 1/\min(T_1,T_2)$ which is the maximum energy graviton she can radiate from accelerating her state.

To obtain an analytic estimate of the decoherence, we can expand the integral in \eqref{eq:N-Minkowski} around small frequencies which gives the dominant contribution to the decoherence. The oscillatory term that appears in $\hat{d^2}(\omega)$ is problematic when expanding around small frequencies, since each higher order in the expansion contains larger powers of $T$, giving non-negligible contributions compared to the lowest-order term for large $T$. With $\sin(\omega T/2) \lesssim 1$, we approximate
\begin{equation}
    d^2(\omega) \sim d_0^2/\omega, \label{eq:dsqFT-order-of-Wald}
\end{equation} 
to obtain an `order-of' estimate for the decoherence of Alice's state. The integrand of \eqref{eq:N-Minkowski} is then (up to a constant) just the mode function multiplied by some power of $\omega$ for each angular momentum mode $l$. For $\omega r \ll 1$, the $l=2$ mode dominates, so to leading order,
\begin{equation}
    \cev{R}_{\omega 2}(r) = \frac{2}{15} \rmi r^3 \omega ^3 + \mathcal{O}(\omega^5). \label{eq:R-soft}
\end{equation}
The number of gravitons Alice radiates is therefore
\begin{equation}
    \avg{N} \sim \frac{m^2 d_0^4}{\min (T_1,T_2)^4}. \label{eq:Wald-decoherence-GR}
\end{equation}
Although it was not explicitly shown in \cite{local-decoherence}, the NLO correction to \eqref{eq:Wald-decoherence-GR}, corresponding to the $\mathcal{O}(\omega^5)$ contribution in \eqref{eq:R-soft}, is $m^2 d_0^4 r^2/\min(T_1,T_2)^6$. This will be useful in the analysis of our calculations in section \ref{ch:GW-decoherence-QFT}. The decoherence \eqref{eq:Wald-decoherence-GR} only depends on the duration of the creation and recombination processes where her state deviates from an inertial trajectory and is accelerated by the Stern-Gerlach apparatus, causing it to emit gravitational radiation.


\section{Decoherence due to gravitational waves} \label{ch:decoherence_gws}
\begin{figure}
\centering
 \begin{subfigure}[b]{0.48\textwidth}
     \centering
    \includegraphics[width=\linewidth]{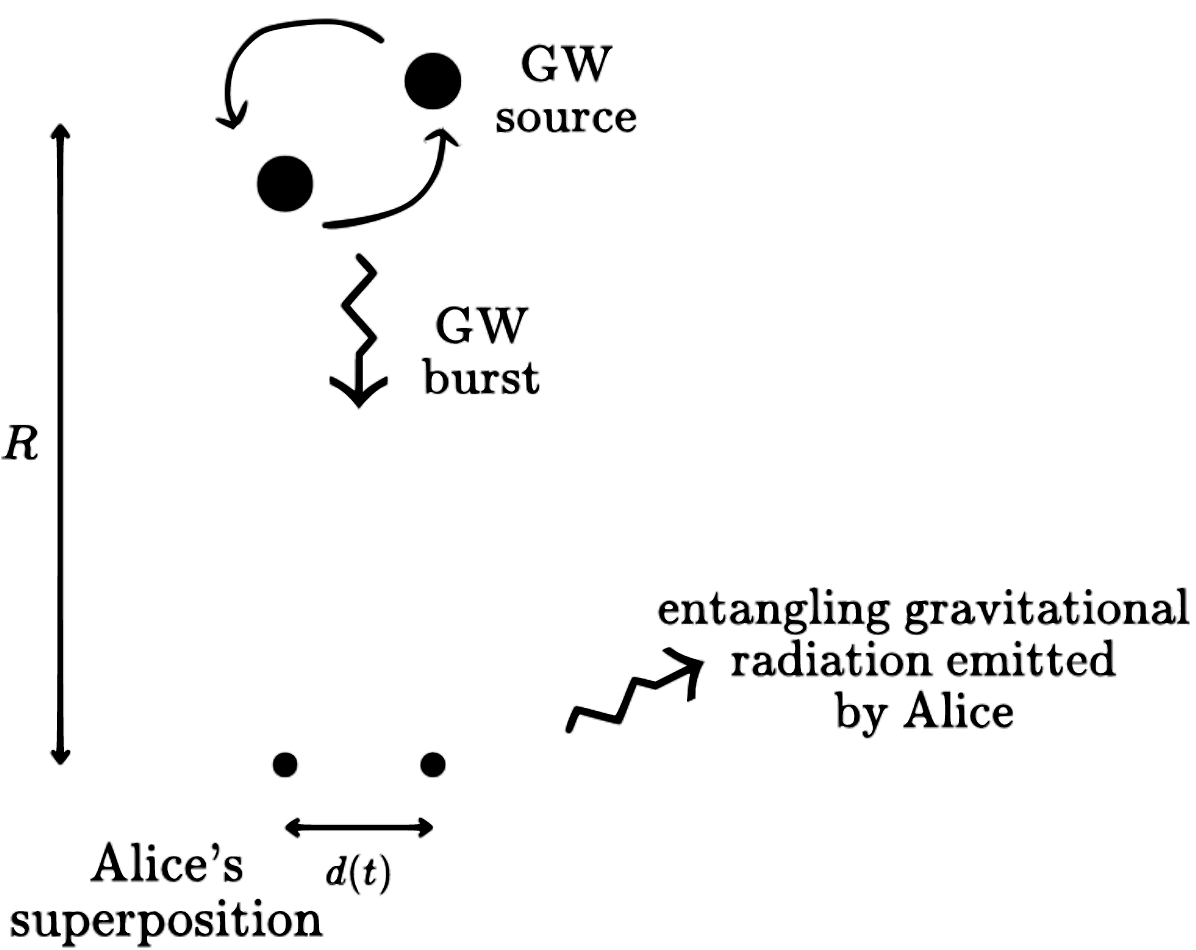}
    \caption{}
    \label{fig:GW-decoherence-setup}
 \end{subfigure}
 \hfill
 \begin{subfigure}[b]{0.48\textwidth}
     \centering
    \includegraphics[width=\linewidth]{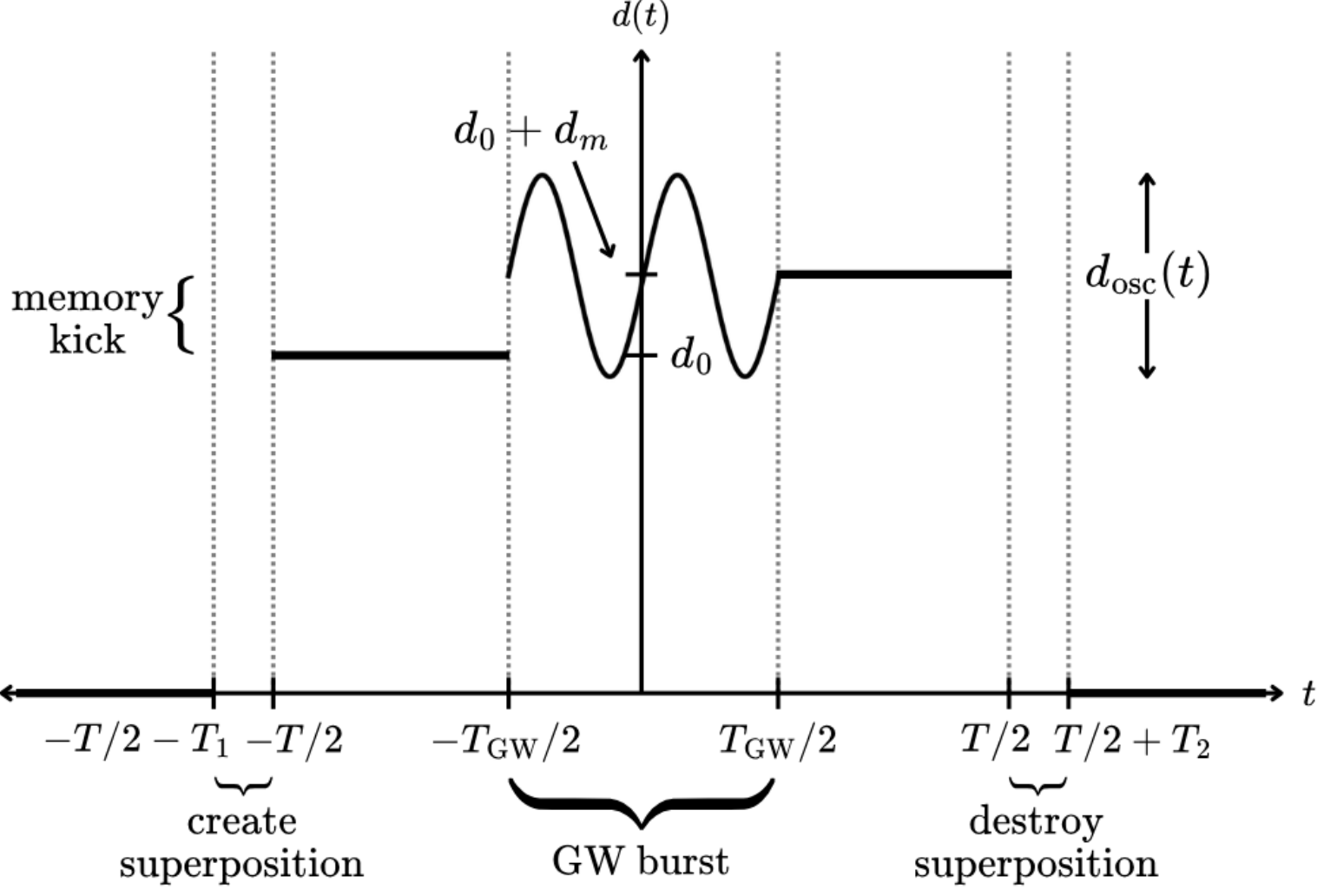}
    \caption{}
    \label{fig:d(t)}
 \end{subfigure}
    \caption{The setup of the modified gedankenexperiment we consider in this paper. (a) A gravitational wave source at distance $R \gg d(t)$ from Alice emits a burst of gravitational radiation incident on her state which generates a time-dependent gravitational quadrupole, causing her to emit gravitational radiation. (b) Displacement between the components of Alice's state for the duration of her experiment. Alice creates the superposition over time $T_1$, initially separating the components by $d_0$ using a Stern-Gerlach device. While she maintains her superposition, she is subject to a gravitational wave burst over time $T_{\text{GW}}<T$, causing her state to oscillate according to $d_{\text{osc}}(t)$ and acquire a permanent displacement $d_{\text{mem}}(t)$ due to memory. After time $T$, she recombines her superposition in time $T_2$ using a reverse Stern-Gerlach device and measures some property of her state.}
    \label{fig:modified-gedankenexperiment}
\end{figure}

In this section, we consider the modified gedankenexperiment setup described in Figure \ref{fig:modified-gedankenexperiment}. After Alice creates her superposition, a gravitational wave burst lasting time $T_{\text{GW}}$ is incident on her state. This gives rise to a time-dependent displacement between the components of her superposition. We consider a simple oscillatory waveform which displaces the components of her state by
\begin{equation}
    d_{\text{osc}}(t) = (d_0/2) h \sin \left( \omega_0 (t+T_{\text{GW}}/2) \right), \label{eq:dosc}
\end{equation}
where $h \ll 1$ is the shear of the monochromatic gravitational wave source with frequency $\omega_0$. We introduce the phase term such that the displacement is zero when the gravitational wave reaches Alice at $t=-T_{\text{GW}}/2$. The components of her superposition will also be permanently displaced due to the non-linear memory carried by the incident gravitational wave due to the emission of soft gravitons to null infinity. This is the same memory effect that appears in the work of Danielson \textit{et al} \cite{danielson_gravitationally_2022,danielson-decoherence1,danielson-decoherence2,local-decoherence}. For a short burst, we suppose that the displacement due to the memory waveform can be approximated by a simple step-function,
\begin{equation}
    d_{\text{mem}}(t) = d_\text{m} \Theta(t+T_{\text{GW}}/2), \label{eq:dmem}
\end{equation}
such that the components of Alice's state are permanently displaced by $d_\text{m}$ for $t>-T_{\text{GW}}/2$. During the burst, the displacement between components of her state is therefore $d(t) = d_0 + d_{\text{mem}}(t) + d_{\text{osc}}(t)$. After the burst, the memory effect persists, and the difference in the phase of the gravitational wave between $t=-T_{\text{GW}}/2$ and $t=T_{\text{GW}}/2$ gives a small permanent displacement, so $d(t)=d_0+d_\text{m}+d_{\text{osc}}(T_{\text{GW}}/2)$ until she recombines her state. The displacement of Alice for the duration of her experiment is shown in Figure \ref{fig:d(t)}.

First, we show that the oscillatory motion of the components of Alice's state due to the incoming gravitational wave gives rise to gravitational quadrupole radiation being emitted and therefore causes decoherence which is proportional to the gravitational memory. Next, we calculate the decoherence using the local quantum field theory description of Alice's gravitational field, originally developed by Danielson \textit{et al} \cite{local-decoherence} and reviewed in section \ref{ch:local-decoherence}, and show that these results are equivalent at leading order. We then compare the decoherence calculation with gravitational wave switched on/off to isolate the decoherence due to the gravitational wave. We will follow a similar procedure to the purely Minkowski case in the previous section. In all `order-of' estimates given, we discard any numeric constants and all final results are truncated to $\mathcal{O}(h,d_\text{m})$.
%


    \subsection{Decoherence due to a gravitational quadrupole} \label{ch:decoherence_quad}
    In the previous section, the creation and recombination of Alice's state led to a time-dependence in the separation $d=d(t)$ between components of her superposition. This is also apparent during the  gravitational wave burst due to both its oscillatory and memory components. Alice's state then has a time-dependent quadrupole moment $Q = m d^2$ during her experiment, which will radiate gravitational energy
    \begin{equation}
        E_{\text{rad}} = \int \rmd t \, P(t) = \frac{2}{5} \int \rmd t \, |\dddot{Q}_{ij}|^2, \label{eq:Erad}
    \end{equation}
    where $P = (2/5)|\dddot{Q}_{ij}|^2$ is the power of the dominant (quadrupole) radiative mode of the gravitational field. Taking $E_{\text{grav}}$ to be the energy of a single graviton, Alice radiates $\avg{N} = E_{\text{rad}}/E_{\text{grav}}$ gravitons. Consequently, her quadrupole will cause her state to radiate \textit{its own} gravitational waves according to
    \begin{equation}
        \dddot{Q}_{ij} = \frac{r}{2} \dot{h}^{\text{A}}_{ij}, \label{eq:quad-Alice}
    \end{equation}
    where the superscript `A' denotes the gravitational wave emitted by Alice's state in the transverse-traceless gauge. Alice's decoherence is thus determined entirely by $h_{ij}^{\text{A}}$. Using \eqref{eq:quad-Alice}, we obtain
    \begin{equation}
        \avg{N} = \frac{1}{E_{\text{grav}}} \frac{r^2}{10} \int \rmd t \, |\dot{h}_{ij}^{\text{A}}|^2, \label{eq:holy-grail-of-my-thesis}
    \end{equation}
    where the bounds extend over the duration of Alice's experiment. Before and after her experiment, Alice's quadrupole vanishes so the bounds of the integral in the above expression can be set to $\pm \infty$. Therefore, the integral in \eqref{eq:holy-grail-of-my-thesis} is exactly the memory carried by the gravitational waves emitted by Alice, reinforcing the idea that the decoherence she experiences is dominated by soft gravitons and is a long-wavelength effect. Remarkably, if an observer were to measure these gravitational waves, they would be able to determine how much her state had decohered! Although \eqref{eq:holy-grail-of-my-thesis} provides a wonderful link between gravitational memory and decoherence of a quantum state, it is far more convenient to calculate the quadrupole moment to estimate the decoherence time. Alternatively, one must determine $h_{ij}^{\text{A}}$ explicitly or measure the amplitude and frequency of the gravitational waves emitted by Alice to obtain $h_{ij}^{\text{A}}$.
     
    We will split up the calculation of the decoherence into the three parts where her gravitational quadrupole radiates: (i) The creation process; (ii) the recombination process; and (iii) during the gravitational wave burst. Alice creates her superposition, whose components are separated by $d_0$, over time $T_1$, so the lowest-energy graviton she can emit has energy $E_{\text{grav}} \sim 1 / T_1$. Following the calculation of Danielson \textit{et al} \cite{danielson_gravitationally_2022}, we approximate the integral in \eqref{eq:Erad} by
    \begin{equation}
        \int \rmd t \, |\dddot{Q}_{ij}|^2 \sim \int \rmd t \, \left( \frac{m d_0^2}{T_1^3} \right)^2 \sim \frac{m^2 d_0^4}{T_1^5}.
    \end{equation}
    The minimum number of gravitons she radiates is therefore
    \begin{equation}
        \avg{N}_{\text{c}} \sim \frac{m^2 d_0^4}{T_1^4}. \label{eq:quad-decoherence-1}
    \end{equation}
    The components of Alice's superposition will be permanently displaced $d_0+d_\text{m}+d_{\text{osc}}(T_{\text{GW}}/2)$ when she begins recombination over time $T_2$. The number of gravitons she radiates is therefore equivalent to \eqref{eq:quad-decoherence-1}, with the replacements $d_0 \to d_0 + d_\text{m} + d_{\text{osc}}(T_{\text{GW}}/2)$ and $T_1 \to T_2$. Keeping only the lowest-order terms (those linear in $h$ and $d_\text{m}$), we have
    \begin{equation}
        \avg{N}_{\text{r}} \sim \frac{m^2 \left( d_0^4 + d_0^3 d_\text{m} + d_0^4 h \sin(\omega_0 T_{\text{GW}}) \right)}{T_2^4}. \label{eq:quad-decoherence-2}
    \end{equation}
    The form of $d(t)$ is more complicated (and oscillatory) during the burst, so we explicitly compute the integral over the quadrupole moment rather than follow the `order-of' estimates of \eqref{eq:quad-decoherence-1} and \eqref{eq:quad-decoherence-2}. The burst lasts for time $T_{\text{GW}}$, so the lowest energy graviton Alice can emit is $E_{\text{grav}} \sim 1/T_{\text{GW}}$. Inserting $Q(t) = m d(t)^2$, we find
    \begin{align}
        \avg{N}_{\text{GW}} &\sim m^2 (1/T_{\text{GW}})^{-1} \int_{-T_{\text{GW}}/2}^{T_{\text{GW}}/2} \rmd t \, (3\dot{d} \ddot{d} + d \dddot{d})^2 \nonumber \\
        %
        &\sim m^2 d_0^4 \omega_0^6 T_{\text{GW}}^2 h^2 \left( 1 + \frac{\sin(\omega_0 T_{\text{GW}})}{\omega_0 T_{\text{GW}}} \right) + \mathcal{O}(h^3, d_\text{m} h^2) \nonumber \\
        &\sim m^2 d_0^4 \omega_0^6 T_{\text{GW}}^2 h^2. \label{eq:quad-decoherence-GW}
    \end{align}
    In the last line, we have used that $1 + \sin{x}/x \sim 1$. The decoherence from \eqref{eq:quad-decoherence-GW} is inherently due to the oscillatory component of the incident gravitational wave and is an $\mathcal{O}(h^2)$ effect, so it is highly suppressed. This can be seen by taking the limit as the source frequency $\omega_0$ tends to zero, for which $\avg{N}_{\text{GW}}$ vanishes. Higher-order multipole moments will also contribute to the hard spectra, however, these are sub-leading effects which are further suppressed by higher powers of the gravitational shear. The total number of gravitons emitted during Alice's experiment is therefore
    \begin{align}
        \avg{N} &= \avg{N}_{\text{c}} + \avg{N}_{\text{r}} + \mathcal{O}(h^2, d_\text{m}^2, d_\text{m} h) 
    \end{align}
    Suppose Alice creates and recombines her state such that $T_1 \sim T_2 \coloneqq T_{\text{SG}}$. Therefore,
    \begin{equation}
        \avg{N} \sim \frac{m^2 \left( d_0^4 + d_0^3 d_\text{m} + d_0^4 h \sin(\omega_0 T_{\text{GW}}) \right)}{T_{\text{SG}}^4}, \label{eq:N-quad}
    \end{equation}
    so a burst of pure memory will decohere Alice's state at leading order and the oscillatory component only contributes due to the phase of the gravitational wave when it is switched off at $t=T_{\text{GW}}/2$. Alice's quadrupole is her dominant radiative multipole, so we expect to recover this result when performing the same calculation using the local (quantum field theory) formalism from the previous section.

    \subsection{Gravitational wave decoherence in quantised gravity} \label{ch:GW-decoherence-QFT}
    We wish to determine an `order-of' estimate for the number of gravitons emitted by Alice using the local quantum field theory approach to construct her gravitational field. In a similar manner to section \ref{ch:local-decoherence}, we want to evaluate
    \begin{equation}
        \avg{N} \sim m^2 \sum_{l=2}^{\infty} \frac{1}{r^6} \int_{\omega_{\text{IR}}}^{\omega_{\text{UV}}} \frac{\rmd \omega}{\omega} |\hat{d^2}(\omega)|^2 |\cev{R}_{\omega l}(r)|^2, \label{eq:N-decoherence-2}
    \end{equation}
    although with a more complicated form for $d(t)$, as described in Figure \ref{fig:d(t)}, which includes the effects of an incident gravitational wave. We have imposed the IR and UV frequency cutoffs $\omega_{\text{IR}} \sim 1/T$ and $\omega_{\text{UV}} \sim 1/\min(T_1,T_2)$, which are unchanged from the purely Minkowski case. Our first step is to evaluate the Fourier transform $\hat{d^2}(\omega)$. We begin by writing the Fourier integral as
    \begin{align}
        \hat{d^2}(\omega) &= \left( \int_{-T/2}^{-T_{\text{GW}}/2} + \int_{-T_{\text{GW}}/2}^{T_{\text{GW}}/2} + \int_{T_{\text{GW}}/2}^{T/2} \right) \rmd t \, \rme ^{\rmi\omega t} d^2(t) \nonumber \\
        &= \hat{d_1^2}(\omega) + \hat{d_2^2}(\omega) + \hat{d_3^2}(\omega) \label{eq:dsqFT-2}.
    \end{align}
    Here, we have used that $d(t)=0$ for $t<-T/2-T_1$ and $t>T/2+T_2$. Furthermore, given the hierarchy of timescales we impose ($T \gg T_{\text{GW}} \gg T_1,T_2$), the contribution to $\hat{d^2}(\omega)$ during the creation and recombination times can be made arbitrarily small, so they can be neglected. For $T_{\text{GW}} \gg T_1, T_2$, the part of the integral where the gravitational wave is present has a non-trivial contribution. In \eqref{eq:dsqFT-2}, $\hat{d_1^2}(\omega)$ and $\hat{d_3^2}(\omega)$ correspond to the parts of the Fourier transform before ($-T/2 < t < -T_{\text{GW}}/2$) and after ($T_{\text{GW}}/2 < t < T/2$) the burst, respectively. The final component, $\hat{d_2^2}(\omega)$, describes the contribution during the burst ($-T_{\text{GW}}/2 < t < T_{\text{GW}}/2$).
    
    We recover the purely Minkowski result \eqref{eq:dsqFT-order-of-Wald} when evaluating \eqref{eq:dsqFT-2} with the gravitational wave switched off, by taking the limits as $\omega_0,h,d_\text{m} \to 0$. With the gravitational wave switched on, we still find that the Fourier transform is dominated by soft modes. In particular, $\hat{d_1^2}(\omega)$ and $\hat{d_3^2}(\omega)$ are the Fourier transforms of constants, which are proportional to $1/\omega$, so soft modes dominate. The second term, $\hat{d_2^2}(\omega)$, contains the Fourier transform of a sinusoid, which gives an oscillatory function centred around the source frequency $\omega_0$. For $\omega_0 \gtrsim 1$, this is a `hard' frequency contribution. However, since the oscillation is damped by the gravitational wave amplitude, the hard contribution is highly suppressed and soft modes still dominate.

    
    Before performing the calculation of the decoherence, let us first briefly comment on the relative contributions of the oscillatory and memory parts of the incident gravitational wave on $\hat{d^2}(\omega)$. Switching off the oscillatory part of the gravitational wave, by taking the limit as $\omega_0 \to 0$, we are left with a burst of pure memory. We find,
    \begin{align}
        |\hat{d^2}(\omega)|^2 = &\frac{4 \sin^2(\omega T/2)}{\omega^2} \times \nonumber \\
        &\left[ d_0^4 + 2 d_0^3 d_\text{m} \left( 1 + \frac{\sin(\omega T_{\text{GW}}/2)}{\sin(\omega T/2)} \right) \right]. \qquad \text{(purely memory GW)}
    \end{align}
    Taking the limit as $T_{\text{GW}} \to T$, we recover exactly the result found in \cite{local-decoherence}, with $d_0$ replaced by $d_0 + d_\text{m}$, to linear order in $d_\text{m}$. Therefore, a burst of memory incident to Alice's state immediately as her superposition is created is equivalent to her superposition having an initial separation of $d_0+d_\text{m}$, as one would expect. Conversely, eliminating the memory part of the gravitational wave by taking $d_\text{m} \to 0$ leaves only the oscillatory component. Since $d_{\text{osc}}(t) \sim \mathcal{O}(h)$, the only purely oscillatory term that appears at this order (with memory switched off) is a phase term. Namely, the only $\mathcal{O}(h)$ contribution is proportional to the difference in phase of the gravitational wave over the duration of the burst. We have,
    \begin{align}
        |\hat{d^2}(\omega)|^2 =& \frac{4 d_0^4 \sin^2(\omega T/2)}{\omega^2} + \nonumber \\
        &\frac{4 d_0^4 h \sin(\omega T/2) \sin(\omega_0 T_{\text{GW}})}{\omega^2} \left( \sin(\omega T/2) + \frac{\omega_0^2}{\omega^2 - \omega_0^2} \sin(\omega T_{\text{GW}}/2) \right. \nonumber \\
        &\left. - \frac{2 \omega_0 \omega}{\omega^2 - \omega_0^2} \cos(\omega T_{\text{GW}}/2) \tan(\omega_0 T_{\text{GW}}/2) \right), \qquad \text{(purely oscillatory GW)} \label{eq:FT-osc}
    \end{align}
    where the leading order $\mathcal{O}(1)$ term gives exactly the result one would obtain by switching the gravitational wave off entirely. Therefore, for a purely oscillatory gravitational wave, Alice will only decohere up to a phase which is in agreement with our quadrupole calculation \eqref{eq:N-quad}.
    
\begin{figure}
    \centering
    \begin{subfigure}[b]{0.49\textwidth}
        \centering
        \includegraphics[width=\linewidth]{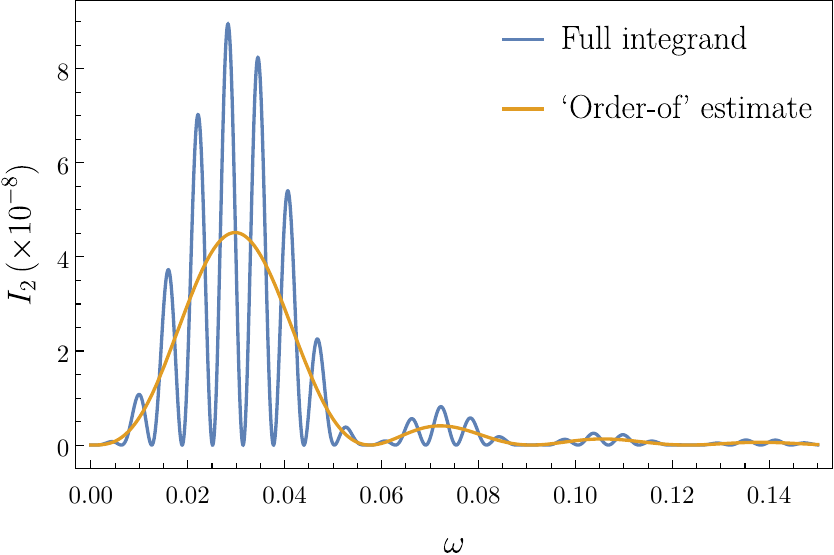}
        \caption{}
        \label{fig:NIntegrand}
    \end{subfigure}%
    \hfill
    \begin{subfigure}[b]{0.49\textwidth}
        \centering
        \includegraphics[width=\linewidth]{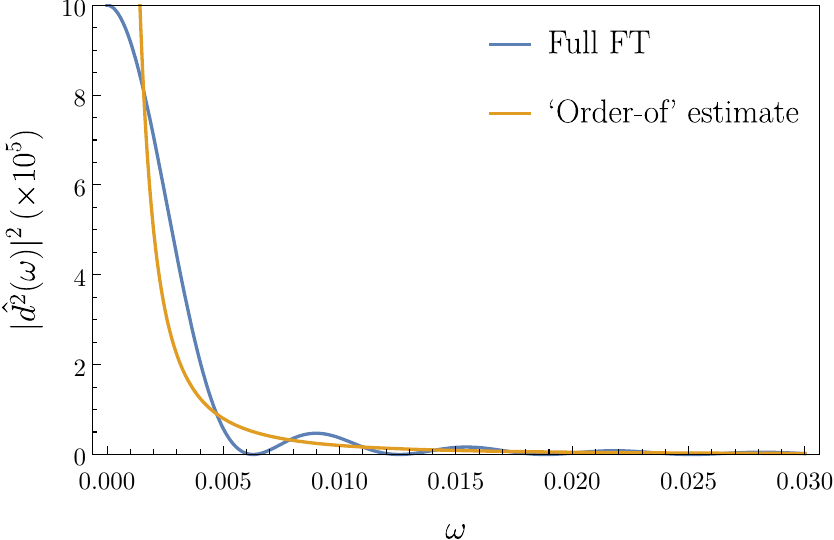}
        \caption{}
        \label{fig:d4FT-comparison}
    \end{subfigure}%
    \caption{(a) $l=2$ mode of the integrand of $\avg{N}$. Integrating over positive frequency modes ($\omega>0$) gives the leading-order contribution to the number of gravitons radiated by Alice during her experiment. The `order-of' integrand displays the same low-frequency behaviour as the full integrand without the approximation. Frequency modes with $\omega \ll 1$ dominate the integral in both cases. (b) Comparison of the full Fourier transform $|\hat{d^2}(\omega)|^2$ and the `order-of' estimate taken by approximating $\sin(\omega T/2) \sim 0$ and $\sin^2(\omega T/2) \sim 1/2$. The `order-of' estimate changes the infrared structure of the Fourier transform by introducing a pole at $\omega = 0$. This divergent behaviour is unproblematic with an appropriate low-frequency cutoff $\omega_{\text{IR}} > 0$. The values used here only serve to demonstrate the generic behaviour of the integrand of $\avg{N}$ and the Fourier transform $\hat{d^2}(\omega)$.}
    \end{figure}

    Next, we expand the integrand around dominant frequencies to obtain an analytic approximation of the decoherence. We define the integrand of $\avg{N}$ contributing to the $l^{\text{th}}$ angular momentum mode as $I_l$, such that
    \begin{equation}
        \avg{N} \sim \sum_{l=2}^{\infty} \int_{\omega_{\text{IR}}}^{\omega_{\text{UV}}} \rmd \omega \, I_l. \label{eq:N-decomposition}
    \end{equation}
    Figure \ref{fig:NIntegrand} shows that the integrand of \eqref{eq:N-decoherence-2} is dominated by soft frequencies, so it is appropriate to expand $I_l$ for $\omega \ll 1$. Once again, we encounter the issue of large powers of $T$ appearing in higher-order terms when expanding the oscillatory terms $\sin(\omega T/2)$ in the Fourier transform, which have non-negligible contributions to the decoherence for $T \gg 1$. Instead, we make an `order-of' estimate by taking the average value of these oscillatory terms. Specifically, we approximate 
    \begin{equation}
    \begin{aligned}
         \sin^2(\omega T/2) &\sim \avg{\sin^2(\omega T/2)} = 1/2, \\
         \sin(\omega T/2) &\sim \avg{\sin(\omega T/2)} = 0,
    \end{aligned}
    \end{equation}
    such that
    \begin{equation}
        |\hat{d^2}(\omega)|^2 \sim \frac{1}{\omega^2} \left( d_0^4 + d_0^3 d_\text{m} + d_0^4 h \sin(\omega_0 T_3) \right).
    \end{equation}
    This differs from the `order-of' estimate used in section \ref{ch:local-decoherence}\footnote{See discussion around \eqref{eq:dsqFT-order-of-Wald}.} since we now have a term linear in $\sin(\omega T/2)$ which is not strictly positive. Using the same approximation as in \eqref{eq:dsqFT-order-of-Wald} would therefore overestimate the number of gravitons Alice radiates. As before, taking such an approximation alters the infrared structure of the integral and introduces a pole at $\omega = 0$, which is regularised by the IR cutoff so $\avg{N}$ remains finite. Figure \ref{fig:d4FT-comparison} shows that the low-frequency behaviour of our `order-of' estimate of the Fourier transform is preserved for $\omega > \omega_{\text{IR}}$, so we expand $I_l$ for small frequencies:
    \begin{equation}
        I_l \sim \frac{m^2}{\omega r^4} (\omega r)^{2l} \left( d_0^4 + d_0^3 d_\text{m} + d_0^4 h \sin(\omega_0 T_{\text{GW}}) \right) \left( 1 - \omega^2 r^2 + \mathcal{O}(\omega^4 r^4) \right).
    \end{equation}
    Since $T_{\text{GW}}$ only appears inside oscillatory terms, we do not need to impose additional bounds on the burst duration for the series to converge well and Alice is free to keep her superposition in-tact for an arbitrary amount of time after the burst hits her state, so long as $T \gg T_{\text{GW}}$ when she recombines and performs her measurement\footnote{Alice will, of course, have to be rather clever to develop a way to keep her superposition in-tact for an appropriate amount of time.}.
    
    For $\omega r \ll 1$, the dominant mode is $l=2$. We find, for $T \gg T_1,T_2$,
    \begin{align}
        \avg{N} \sim \, & m^2 \left( d_0^4 + d_0^3 d_\text{m} + d_0^4 h \sin(\omega_0 T_{\text{GW}}) \right) \left( \frac{1}{\min(T_1,T_2)^4} - \frac{r^2}{\min(T_1,T_2)^6} \right). \label{eq:N-order-of}
    \end{align}
    In the limit as the gravitational wave is turned off, we recover the purely Minkowski result \eqref{eq:Wald-decoherence-GR}. Our expression \eqref{eq:N-order-of} is highly oscillatory in the source frequency and burst duration, so Alice's decoherence is highly dependent on the phase of the gravitational waveform when the burst is switched off.  
    
    The leading order term, proportional to $1/\min(T_1,T_2)^4$ in \eqref{eq:N-order-of}, is exactly the decoherence from Alice's mass quadrupole \eqref{eq:N-quad}, which is dominated by soft graviton emission. The NLO term, which is proportional to $1/\min(T_1,T_2)^6$, is absent from the multipole calculation since it appears as a relic from the frequency expansion used in $I_l$. In \eqref{eq:quad-decoherence-GW}, we showed that \textit{during} the burst, the gravitational wave contributes to the decoherence at $\mathcal{O}(h^2)$. The sub-leading multipoles for the gravitational field therefore provide corrections to $\mathcal{O}(h^2)$ at most\footnote{The subleading radiative multipoles are the mass octupole (which vanishes given the configuration of Alice's experiment) and the current quadrupole (which contributes to $\mathcal{O}(h^2)$) \cite{roskill_mass_2024}.}, so do not contribute to the NLO term here.
    
    To gauge the degree to which the gravitational wave decoheres Alice, we consider the ratio of $\avg{N}$ with the gravitational wave source switched on/off. We find,
    \begin{align}
        \avg{N}/\avg{N}_{\text{(GW off)}} \sim \left(1 + \frac{d_\text{m}}{d_0} + h \sin(\omega_0 T_{\text{GW}}) \right) \left( 1 - \frac{r^2}{\min(T_1,T_2)^2} \right). \label{eq:N-ratio}
    \end{align}
    There is a clear distinction between the memory and oscillatory contributions to the decoherence in \eqref{eq:N-ratio}. In particular, the memory contributions depend on the ratio $d_\text{m}/d_0$ and the oscillatory contributions are suppressed by the gravitational wave's amplitude and its phase when the burst stops. Using the step-function memory model, a larger memory `kick' accelerates her state more rapidly, causing her to decohere further. This is in agreement with Belenchia \textit{et al} \cite{belenchia_quantum_2018}, who showed that if Alice accelerates her state in a non-uniform manner, she will radiate gravitons entangled to her state. Since Alice is accelerated non-uniformly by the oscillatory waveform during the burst, one would expect a longer burst duration $T_{\text{GW}}$ to further decohere her state. However, as suggested by our quadrupole calculation, this decoherence only appears as an $\mathcal{O}(h^2)$ effect so it is highly suppressed. For carefully chosen source parameters, namely, $\omega_0 T_{\text{GW}} \approx n \pi$ ($n\in \mathbb{N}$), the oscillatory contributions nearly vanish, leaving memory as the dominant source of decoherence from the gravitational wave.


\section{Conclusions}
In this work, we quantified the decoherence of a quantum spatial superposition due to an incident gravitational wave burst. In section \ref{ch:decoherence_quad}, we showed that decoherence due to the dominant (quadrupole) radiative mode appears only from the creation and recombination of the superposition at leading order $\mathcal{O}(h,d_\text{m})$. The effect of the incident gravitational wave appears only due to the permanent displacement from memory and the phase of the wave when it is switched off.

We extended this analysis in section \ref{ch:GW-decoherence-QFT} by quantising the gravitational field of the system, incorporating an external wave burst via simplified oscillatory and memory waveforms. Although the burst induces hard-frequency contributions to the decoherence, these are suppressed by the gravitational wave shear and found to be subdominant to soft graviton radiation. For carefully chosen gravitational burst frequencies and durations, the oscillatory contributions to the decoherence nearly vanish entirely, so the memory effect dominates.

To leading order, the decoherence found by quantising the state's gravitational field matched the multipole calculation, which supports the validity of the semi-classical approach used. However, the NLO term, which was absent in the purely classical calculation, appeared naturally in the quantum treatment and may be interpreted as a quantum correction to the decoherence.

This work demonstrates that classical gravitational radiation - particularly the memory effect - limits an experimenter's ability to maintain a coherent quantum superposition. While we restricted our analysis to simplistic oscillatory and memory waveforms to obtain an analytic estimate of the decoherence, this work serves as a step towards understanding how quantum systems interact with classical gravitational fields. More accurate predictions of the decoherence of quantum spatial superpositions due to gravitational waves can be obtained numerically using realistic waveforms produced by typical astrophysical sources to deepen our understanding of how gravity influences quantum systems.

\section*{Acknowledgements}
S. T. is supported by the Swiss National Science Foundation Ambizione Grant Number: PZ00P2-202204.


\section*{Appendix}
\appendix
\setcounter{section}{1}

The memory effect that appears in general relativity is understood to manifest in other gauge theories including electromangetism \cite{color_mem_Pate,Bieri_2012,susskind2019electromagneticmemory,bieri_electromagnetic_2013,pasterski_asymptotic_2017}. A natural analogue of the gedankexperiment we considered is to replace Alice's massive superposition with a superposition of charges $q$ and the gravitational wave burst with a burst of electromagnetic radiation with frequency $\omega_0$ and amplitude $E \ll 1$ for time $T_{\text{EM}} \ll T$. Here, we briefly comment on Alice's decoherence in the electromagnetic analogue. 

Similar to the gravitational case, the wave will cause the components of her superposition to oscillate and acquire a permanent displacement due to memory \cite{bieri_electromagnetic_2013}. The displacement is given by $d(t)$ as described in Figure \ref{fig:d(t)}, with $h \to E$ and $T_{\text{GW}} \to T_{\text{EM}}$. Alice's effective dipole moment $D = q d$ will cause her to emit electromagnetic radiation with power $P \sim |\ddot{D}|^2$. The dipole radiated power during the electromagnetic burst is an $\mathcal{O}(E^2)$ contribution, similar to \eqref{eq:quad-decoherence-GW}, so only the creation and recombination processes contribute at linear order. Due to the difference in the dominant radiative multipole moment between the electromagnetic and gravitational cases, the gravitational decoherence depends more strongly on the frequency of the burst. The total number of photons Alice emits during her experiment is
\begin{equation}
    \avg{N}^{\text{EM}} \sim \frac{q^2 \left( d_0^2 + d_0 d_\text{m} + d_0^2 E \sin(\omega_0 T_{\text{EM}}) \right)}{T_{\text{SG}}^4} + \mathcal{O}(E^2, d_\text{m}^2, E d_\text{m}), \label{eq:N-dipole}
\end{equation}
with $T_1 \sim T_2 \coloneqq T_{\text{SG}}$.

Quantising Alice's electromagnetic field using the local formalism in Minkowski spacetime, the number of photons Alice radiates during her experiment is \cite{local-decoherence}
\begin{equation}
    \avg{N}^{\text{EM}} \sim q^2 \sum_{l=1}^{\infty} \frac{1}{r^4} \int_{0}^{\infty} \frac{\rmd \omega}{\omega} |\hat{d}(\omega)|^2 |\cev{R}_{\omega l}(r)|^2, \label{eq:N-decoherence-EM}
\end{equation}
where
\begin{equation}
    \hat{d}(\omega) = \int \rmd t \, d(t) \rme ^{\rmi\omega t}
\end{equation}
is the Fourier transform of $d(t)$ and the mode functions are the same as in the gravitational case.

The IR and UV frequency cutoffs remain unchanged in the electromagnetic case. Once again, soft modes dominate the Fourier transform $\hat{d}(\omega)$, so we expand the integrand for $\omega \ll 1$. We perform a similar `order-of' estimate for the Fourier transform by approximating oscillatory terms involving $T$ by their average value. Unlike the gravitational case, the leading angular momentum mode is for $l=1$ with $\omega r \ll 1$, so the leading order term in the integrand of $\avg{N}^{\text{EM}}$ is $\mathcal{O}(\omega)$ rather than $\mathcal{O}(\omega^3)$. We find
    \begin{equation}
        \avg{N}^{\text{EM}} \sim q^2 \left( d_0^2 + d_0 d_\text{m} + d_0^2 E \sin(\omega_0 T_{\text{EM}}) \right) \left( \frac{1}{\min(T_1,T_2)^2} - \frac{r^2}{\min(T_1,T_2)^4} \right).
    \end{equation}
    In the limit as the electromagnetic wave source is switched off, we recover decoherence originally found by \cite{danielson_gravitationally_2022}. Similar to the gravitational case, the leading order term is exactly the decoherence due to Alice's dipole radiation \eqref{eq:N-dipole}. With the incoming gravitational/electromagnetic source turned on, there is still a direct correspondence between the two cases. Namely, they are related by $q \to m$, $h \to E$, $T_{\text{GW}} \to T_{\text{EM}}$ and multiplying each term by $d_0^2/\min(T_1,T_2)^2$, which holds even for beyond-leading order contributions at $\mathcal{O}(h,d_\text{m})$\footnote{$\mathcal{O}(E,d_\text{m})$ in the electromagnetic case}.

\printbibliography

\end{document}